\documentclass[12pt,preprint]{aastex62}

\usepackage[T1]{fontenc}
\usepackage{pslatex}
\usepackage{contour}
\usepackage{color}

\usepackage{graphicx}
\usepackage{epsfig} 
\usepackage{natbib}
\usepackage{times}
\usepackage{amssymb,amsmath}
\usepackage{color}
\usepackage{afterpage}
\usepackage{comment}

\definecolor{dark-red}{rgb}{0.8,0,0}
\definecolor{dark-green}{rgb}{0,0.4,0}
\definecolor{dark-blue}{rgb}{0,0,0.8}
\definecolor{dark-margenta}{rgb}{0.8,0,0.8}
\definecolor{orange}{rgb}{1.0,0.6,0}
\definecolor{grey}{rgb}{0.6,0.6,0.6}
\newcommand{\hB}{ }

\newcommand{\mybold}{ }

\newcommand{\stth}[1]{}
\definecolor{cOOO}{RGB}{230, 230, 230}   % 7
\definecolor{cOOD}{RGB}{170,  10, 190}   % 6  -1
\definecolor{cOCO}{RGB}{  0, 140,  70}   % 5  -2
\definecolor{cOCC}{RGB}{190,   0,   0}   % 4  -3
\definecolor{cDOO}{RGB}{  0,   0, 255}   % 3
\definecolor{cDOC}{RGB}{255, 255,  30}   % 2
\definecolor{cDDO}{RGB}{  0, 190, 190}   % 1  -4
\definecolor{cDDD}{RGB}{150, 150, 150}   % 0
\newcommand{\DDD}{\contour{black}{\textcolor{cDDD}{\textbf{DDD}}}}
\newcommand{\DDO}{\contour{black}{\textcolor{cDDO}{\textbf{DDO}}}}
\newcommand{\DOC}{\contour{black}{\textcolor{cDOC}{\textbf{DOC}}}}
\newcommand{\DOO}{\contour{black}{\textcolor{cDOO}{\textbf{DOO}}}}
\newcommand{\OCC}{\contour{black}{\textcolor{cOCC}{\textbf{OCC}}}}
\newcommand{\OCO}{\contour{black}{\textcolor{cOCO}{\textbf{OCO}}}}
\newcommand{\OOD}{\contour{black}{\textcolor{cOOD}{\textbf{OOD}}}}
\definecolor{cOOO}{RGB}{230, 230, 230}   % 7
\definecolor{cOOC}{RGB}{255,  30, 255}   % 6  -1
\definecolor{cODO}{RGB}{  0, 240,   0}   % 5  -2
\definecolor{cODD}{RGB}{  0, 240, 240}   % 4  -3
\definecolor{cCOO}{RGB}{  0,   0, 255}   % 3
\definecolor{cCOD}{RGB}{255, 255,  30}   % 2
\definecolor{cCCO}{RGB}{255,   0,   0}   % 1  -4
\definecolor{cCCC}{RGB}{150, 150, 150}   % 0
\newcommand{\CCC}{\contour{black}{\textcolor{cCCC}{\textbf{CCC}}}}
\newcommand{\CCO}{\contour{black}{\textcolor{cCCO}{\textbf{CCO}}}}
\newcommand{\COD}{\contour{black}{\textcolor{cCOD}{\textbf{COD}}}}
\newcommand{\COO}{\contour{black}{\textcolor{cCOO}{\textbf{COO}}}}
\newcommand{\ODD}{\contour{black}{\textcolor{cODD}{\textbf{ODD}}}}
\newcommand{\ODO}{\contour{black}{\textcolor{cODO}{\textbf{ODO}}}}
\newcommand{\OOC}{\contour{black}{\textcolor{cOOC}{\textbf{OOC}}}}
\newcommand{\OOO}{\contour{black}{\textcolor{cOOO}{\textbf{OOO}}}}

\definecolor{cOOOi}{RGB}{230, 230, 230}   % 7
\definecolor{cOODi}{RGB}{190,  80, 160}   % 6   -1
\definecolor{cOCOi}{RGB}{  0, 140,  70}   % 5   -2
\definecolor{cOCCi}{RGB}{190,   0,   0}   % 4   -3
\definecolor{cDOOi}{RGB}{190,   0,   0}   % 3
\definecolor{cDOCi}{RGB}{255, 255,  30}   % 2
\definecolor{cDDOi}{RGB}{220,  50,   0}   % 1   -4
\definecolor{cDDDi}{RGB}{150, 150, 150}   % 0

\newcommand{\DDDi}{\contour{black}{\textcolor{cDDDi}{\textbf{DDD}}}}
\newcommand{\DDOi}{\contour{black}{\textcolor{cDDOi}{\textbf{DDO}}}}
\newcommand{\DOCi}{\contour{black}{\textcolor{cDOCi}{\textbf{DOC}}}}
\newcommand{\DOOi}{\contour{black}{\textcolor{cDOOi}{\textbf{DOO}}}}
\newcommand{\OCCi}{\contour{black}{\textcolor{cOCCi}{\textbf{OCC}}}}
\newcommand{\OCOi}{\contour{black}{\textcolor{cOCOi}{\textbf{OCO}}}}
\newcommand{\OODi}{\contour{black}{\textcolor{cOODi}{\textbf{OOD}}}}
\definecolor{cOOOi}{RGB}{230, 230, 230}   % 7
\definecolor{cOOCi}{RGB}{255, 128, 128}   % 6  -1
\definecolor{cODOi}{RGB}{  0, 240,   0}   % 5  -2
\definecolor{cODDi}{RGB}{255,   0,   0}   % 4  -3
\definecolor{cCOOi}{RGB}{190,   0,   0}   % 3
\definecolor{cCODi}{RGB}{255, 255,  30}   % 2
\definecolor{cCCOi}{RGB}{190,   0,   0}   % 1  -4
\definecolor{cCCCi}{RGB}{150, 150, 150}   % 0

\newcommand{\CCCi}{\contour{black}{\textcolor{cCCCi}{\textbf{CCC}}}}
\newcommand{\CCOi}{\contour{black}{\textcolor{cCCOi}{\textbf{CCO}}}}
\newcommand{\CODi}{\contour{black}{\textcolor{cCODi}{\textbf{COD}}}}
\newcommand{\COOi}{\contour{black}{\textcolor{cCOOi}{\textbf{COO}}}}
\newcommand{\ODDi}{\contour{black}{\textcolor{cODDi}{\textbf{ODD}}}}
\newcommand{\ODOi}{\contour{black}{\textcolor{cODOi}{\textbf{ODO}}}}
\newcommand{\OOCi}{\contour{black}{\textcolor{cOOCi}{\textbf{OOC}}}}
\newcommand{\OOOi}{\contour{black}{\textcolor{cOOOi}{\textbf{OOO}}}}

\DeclareMathOperator{\slog}{s-log}

\newcommand{\pP}{\mathrm{P}}
\newcommand{\tP}{\mathrm{\tilde{P}}}
\newcommand{\vfp}{\mathbf{v}_{\mathrm{FP}}}
\newcommand{\vtfl}{\mathbf{v}_{\mathrm{TFL}}}

\shortauthors{Lionello et al.}
\shorttitle{Slip-Back Mapping as a Tracker of Topological Changes}

\begin{document}

%\title{Slip-Back Mapping as a Tracker of Topological Changes in MHD Simulations}   
\title{Slip-Back Mapping as a Tracker of Topological Changes in Evolving Magnetic Configurations}   

\correspondingauthor{R.~Lionello}
\email{lionel@predsci.com}

\author{R.~Lionello}
\affiliation{Predictive Science Inc., 9990 Mesa Rim Rd., Ste. 170, San Diego, CA 92121, USA} 

\author{V.~S.~Titov}
\affiliation{Predictive Science Inc., 9990 Mesa Rim Rd., Ste. 170, San Diego, CA 92121, USA} 

\author{Z.~Miki\'c}
\affiliation{Predictive Science Inc., 9990 Mesa Rim Rd., Ste. 170, San Diego, CA 92121, USA} 

\author{J.~A.~Linker}
\affiliation{Predictive Science Inc., 9990 Mesa Rim Rd., Ste. 170, San Diego, CA 92121, USA}

\begin{abstract}
The topology of the coronal magnetic field produces a strong impact on the properties of the solar corona and presumably on the origin of the slow solar wind.
To advance our understanding of this impact,
we revisit the concept of the so-called slip-back mapping \citep{2009ApJ...693.1029T} and adapt it to determine open, closed, and disconnected flux systems that are formed in the solar corona  by magnetic reconnection during a given time interval.
In particular,
the method we developed allows us to describe the magnetic flux transfer between open and closed flux regions via so-called interchange reconnection with {\hB an} unprecedented level of detail. 
We illustrate the application of this method to the analysis of {\hB the} global MHD evolution of the solar corona driven by idealized differential rotation of the photospheric plasma.
\end{abstract}


%%%%%%%%%%%%%%%%%%%%%%
%%%%%%%%%%%%%%%%%%%%%%
\section{Introduction}
\label{s:intro}
%%%%%%%%%%%%%%%%%%%%%%
%%%%%%%%%%%%%%%%%%%%%%
% http://adsabs.harvard.edu/cgi-bin/nph-abs_connect?db_key=AST&qform=PHY&arxiv_sel=astro-ph&arxiv_sel=cond-mat&arxiv_sel=cs&arxiv_sel=gr-qc&arxiv_sel=hep-ex&arxiv_sel=hep-lat&arxiv_sel=hep-ph&arxiv_sel=hep-th&arxiv_sel=math&arxiv_sel=math-ph&arxiv_sel=nlin&arxiv_sel=nucl-ex&arxiv_sel=nucl-th&arxiv_sel=physics&arxiv_sel=quant-ph&arxiv_sel=q-bio&aut_logic=OR&author=&ned_query=YES&sim_query=YES&start_mon=&start_year=&end_mon=&end_year=&ttl_logic=AND&title=solar+wind+topology&txt_logic=AND&text=&nr_to_return=200&start_nr=1&jou_pick=NO&ref_stems=&data_and=ALL&group_and=ALL&start_entry_day=&start_entry_mon=&start_entry_year=&end_entry_day=&end_entry_mon=&end_entry_year=&min_score=&sort=SCORE&data_type=SHORT&aut_syn=YES&ttl_syn=YES&txt_syn=YES&aut_wt=1.0&ttl_wt=0.3&txt_wt=3.0&aut_wgt=YES&obj_wgt=YES&ttl_wgt=YES&txt_wgt=YES&ttl_sco=YES&txt_sco=YES&version=1

Three-dimensional (3D) numerical models, particular magnetohydrodynamic (MHD) simulations, play an increasingly prominent role in attempts to interpret observations of solar and heliospheric phenomena, and to achieve deeper insights into the fundamental processes that drive them.   The role of topological changes in the magnetic field, particularly magnetic reconnection, is often of central interest.  A specific example is the origin of the slow solar wind.  One group of theories \citep[e.g.,][]{2007ApJS..171..520C,2012ApJ...749..182W} assumes that the slow wind primarily arises quasi-statically from regions of large expansion factor near the boundaries of coronal holes, while a contrasting set of theories argue that the slow solar wind is dynamic in origin and involves the reconnection and exchange of {\hB plasma in open- and closed-field regions} \citep{1998SSRv...86...51F,1999ApJ...521..859S,2011ApJ...731..112A}.  While a dynamic component to the slow solar wind is clearly apparent in white light coronagraph \citep{2000JGR...10525133W} and heliospheric images \citep{2010JGRA..115.4103R,2010JGRA..115.4104R}, the question of whether topological changes to the field play a significant role in creating the slow wind, or only a minor one, is controversial.  The S-web model, which was introduced in a series of three papers
\citep{2011ApJ...731..112A,2011ApJ...731..110L,2011ApJ...731..111T} and continues to be investigated \citep{2015ApJ...805...39P,2018ApJ...859....6H}, identifies a set of separatrix surfaces and quasi-separatrix layers (QSLs) in the vicinity of the streamer belt.  These structures provide a generally favorable condition for magnetic reconnection to occur, and are hypothesized to serve as a source of the slow solar wind.  Time-dependent MHD simulations can play an important role in assessing the importance of this process, but only if we can identify and quantify the reconnection in realistic 3D models of the dynamic solar corona.  The complexity of the 3D data that arises from these calculations makes this a seemingly daunting task.

%Fortunately, a general method for describing magnetic reconnection in terms of the variation of the field line connectivity has been already proposed \citep{2009ApJ...693.1029T}. 
%This method utilizes the calculation of the so-called slip-back and slip-forth mappings in evolving magnetic configurations in order to detect differences between the actual field-line topology and that  caused by an ideal evolution within a given time interval.
{\hB Fortunately, the search for a general method to  describe magnetic 
reconnection  has incrementally progressed through the years.
 \citet{1988JGR....93.5547S} and \citet{1988JGR....93.5559H}  showed that the 
three-dimensional
reconnection process develops if the plasma-magnetic configuration  contains a localized nonideal region with an enhanced parallel electric field  $E_{\|}=({\bf E \cdot B})/B$, where ${\bf E}$ and ${\bf B}$ are electric and magnetic fields, respectively.
It was also demonstrated that the rate of reconnection is determined by the maximum of the voltage difference induced by $E_\|$ in this region.
For an MHD configuration with plasma resistivity $\eta$, parallel current density  $j_{\|}$,  and Ohm's law $E_{\|} = \eta j_{\|}$,
this implies  in turn a locally enhanced magnetic ``diffusion region,''
as it was originally named. Clearly the magnetic diffusion that leads to reconnection develops also due to the presence of ${\bf j}_{\perp}$, as it is evident for two-dimensional MHD flows where ${\bf B}\cdot {\bf j} = {\bf E}\cdot {\bf B} =0$ throughout the volume.
In this respect, it appears surprising that the rate of reconnection in a three-dimensional MHD configuration with a nonvanishing ${\bf B}$ in the nonideal region does not depend on ${\bf j}_{\perp}$.
In any case, there is an ongoing slippage in that region between plasma elements and  the magnetic field lines connecting them,
causing plasma that enters the nonideal region to change its magnetic
connection.

\citet{1984GMS....30....1A} used {\hB this change in connectivity} as a characteristic feature to define the reconnection process.  
In an MHD plasma with a finite resistivity, $E_{\|}$ is present everywhere in the volume and it can be much larger in certain locations compared with the remaining volume, either because of  a local enhancement of $j_{\|}$ or 
$\eta$, or both.
The bulk effect of  magnetic diffusion, irrespective of whether it is localized or spread throughout the volume, causes magnetic field lines to change their connections to the boundary, even if the normal component of ${\bf B}$ to it does not change.
The displacements of initially conjugate footpoints relative to their initial positions grow with time and with the voltage difference induced by $E_{\|}$ along the corresponding field lines \citep{2005ApJ...631.1227H}.
If this voltage difference is developed primarily across the nonideal region,
the resulting footpoint displacements would be solely due to the reconnection process occurring in this region.

Based on this premise, \citet{2009ApJ...693.1029T} suggested to quantify reconnection in MHD simulations by measuring such connectivity changes.
For closed-field configurations in the solar corona with no emergence or submergence of the magnetic flux, such changes can be determined by tracking the footpoints of magnetic field lines
via only the plasma velocity field at the boundary and the magnetic field in the volume.
The $E_{\|}$-distribution is in principle not needed for such calculations, but it can be used as an independent method for quantifying reconnection, whose results should be consistent with those derived from the changes in connectivity.

The method of \citet{2009ApJ...693.1029T} employs a composition of four different mappings of the configuration boundary.
Two of them are field-line mappings of the boundary on itself at two different times,  $t_{0}$ and $t_{1}$, of the simulated evolution.
The other two mappings define the footpoint displacements that the corresponding magnetic field lines would experience within the time interval $[t_{0}, t_{1}]$ in an {\it ideal} MHD evolution.
Depending on the order in which these four mappings are composed, one obtains the footpoint displacements that have to occur or have occurred within the time interval $[t_{0}, t_{1}]$ in reference to the configuration at times $t_{0}$ and $t_{1}$, respectively.
The corresponding composite mappings were called slip-mapping forward and backward in time, or briefly  slip-forth and slip-back mappings. 
By analyzing these mappings one can identify the reconnected magnetic flux systems in closed-field configurations whose evolution is driven by horizontal footpoint displacements.}

The primary purpose of this paper is to show that a suitable adaptation of this method to global coronal configurations can provide a wealth of information on how magnetic reconnection develops under coronal conditions.  In particular, the method can quantitatively identify the regions in the modeled solar wind that underwent interchange reconnection \citep[e.g.,][]{2002JGRA.107b.SSH3C}, the process believed to be essential for producing the {\hB unsteady} slow solar wind.

To develop our new method, we have specialized  the concept of slip-back mapping to track the opening, closing, and {\hB disconnection} of coronal field lines in evolving magnetic configurations.
As an illustration of the method, we apply it to the MHD simulation of an evolving solar corona \citep{2005ApJ...625..463L} driven by differential rotation of the Sun with a photospheric magnetic field distribution similar to that of \citet{1996Sci...271..464W}, which is a superposition of bipolar and background dipole fields.
This simulation revealed changes in the magnetic topology and examples of interchange reconnection in the modeled solar corona, which we use here as a benchmark for testing our new method.

This paper is organized as follows: in Section~\ref{sec-SBmap} we describe
the slip-back mapping technique.
Then in Section~\ref{sec-application} we present an application 
of the technique to the analysis of the simulation of \citet{2005ApJ...625..463L}.
We conclude by discussing the relevance of our results.

%%%%%%%%%%%%%%%%%%%%%%
%%%%%%%%%%%%%%%%%%%%%%
\section{SLIP-BACK MAPPING REVISITED}

\label{sec-SBmap}

The slip-back mapping is constructed to identify reconnected magnetic flux systems in evolving magnetic configurations.  In particular, it can be used to track magnetic fluxes {\hB undergoing} interchange reconnection between open and closed magnetic field lines, which may {simultaneously be occurring} at multiple sites in the solar corona.
This process was invoked to explain the properties of the slow solar wind
\citep{1998SSRv...86...51F,2003JGRA..108.1157F,2009IAUS..257..109F}.

\subsection{The Concept and Basic Properties of Slip-Back Mapping}
\label{ssec-concept}

To elucidate this concept, we will use a spherical system of coordinates $(r,\theta,\phi)$ with a time-dependent 3D magnetic field $\mathbf{B}(r,\theta,\phi,t)$,
which can result from an MHD simulation, an analytical model, or
some other form of modeling that provides a magnetic field evolution together with the corresponding plasma flows.
Let $\mathbf{B}$  be defined in the domain between two spheres of radius $r=R_0$ and $r=R_1$ with $R_0<R_1$.
In our simulations of the solar corona, $R_0$ corresponds to the photosphere and equals the radius $R_\sun$ of the Sun, and $R_1$ is typically of the order of $20\, R_\sun$.

One way to track evolving magnetic field lines is to use the surface velocity $\vfp$ of the footpoints as if they were frozen-in and advected with the moving plasma elements.
This velocity field can be expressed in terms of the full velocity $\mathbf{v}$ of the plasma and the magnetic field $\mathbf{B}$ at the boundary as
\begin{eqnarray}
  \vfp =\left.\left( \mathbf{v} - \frac{v_{r}}{B_{r}} \, \mathbf{B} \right) \right|_{r=R_{*}}
 \, ,
	\label{vfp}
\end{eqnarray}
where $R_{*}$ equals either $R_0$ or $R_1$, depending on which spherical boundary is considered.
We exploit here the fact that the subtraction of a component parallel to $\mathbf{B}$ from $\mathbf{v}$ cannot affect the ``frozenness" of plasma elements with the field lines they are threaded by.
The coefficient in front of $\mathbf{B}$ is chosen so as to eliminate the radial velocity component, i.e., the component orthogonal to our spherical boundaries.

It should be noted first that the introduced $\vfp$ does not coincide with the tangential component of $\mathbf{v}$, unless $\mathbf{B}$ is strictly radial at the boundary.
Secondly, if magnetic flux emerges or submerges at the boundary,
so that $v_{r} \ne 0$ at the polarity inversion line (PIL),
the velocity $\vfp$ provided by Equation (\ref{vfp}) becomes singular at the PIL,
where $B_{r}$ vanishes by definition.
This singularity can be envisioned also from the local geometry of magnetic field lines near the PIL.
In general, they are locally of a parabolic shape, so that their footpoints have to diverge from or converge to the PIL with an infinite speed at the instant when the field lines just touch the boundary.
The latter does not contradict physics, since $\vfp$ is only an apparent velocity.
It is not difficult to show that the singularity of $\vfp$ at the PIL is integrable, meaning that the corresponding Lagrangian displacement $\int \vfp(\theta(t),\phi(t),t)\, \mathrm{d} t$ is well defined near the PIL within a certain time interval.
Incorporating this fact into the numerical algorithm for calculating the footpoint displacements, one could, in principle, completely prevent division by zero at the PIL.
However, we find that the same result is technically easier to achieve by simply 
approximating Equation (\ref{vfp}) as
\begin{eqnarray}
  \vfp \approx  \left.\left( \mathbf{v} 
        - \frac{v_{r}\, ( B_{r}\, \mathbf{B} + \epsilon^2 \hat{\mathbf{r}} )}{{B_{r}^2}+\epsilon^2}
  \right) \right|_{r=R_{*}} \, ,
	\label{vfpa}
\end{eqnarray}
where $\epsilon$ is a nonvanishing small parameter.
This expression defines a strictly tangential velocity field, which is regular at the PIL, and which turns into Equation (\ref{vfp}) in the limit of vanishing $\epsilon$.
For a sufficiently small $\epsilon$, Equation (\ref{vfpa}) provides an acceptable approximation of $\vfp$ outside a narrow strip along the PIL, whose width is controlled by $\epsilon$.
We will use it for setting the footpoint velocities $\mathbf{v}_{\mathrm{FP}_0}(\theta,\phi,t)$ and $\mathbf{v}_{\mathrm{FP}_1}(\theta,\phi,t)$ at our spherical boundaries with radius $R_0$ or $R_1$, respectively.
%\begin{verbatim}
%Not at the R0 boundary, where the full velocity is tangential!
%\end{verbatim}
%
%{\hR
%\begin{verbatim}
%This is not a tangential component that is orthogonal to the
%radial component!!!
%This is a velocity component supplementary to the velocity component
%parallel to the local magnetic field (Figure 1).
%In ideal MHD this component would coincide with the surface
%velocity of the field line footpoints.
%\end{verbatim}
%}

\begin{figure*}[t]
\centering
\includegraphics[width=0.9\textwidth]{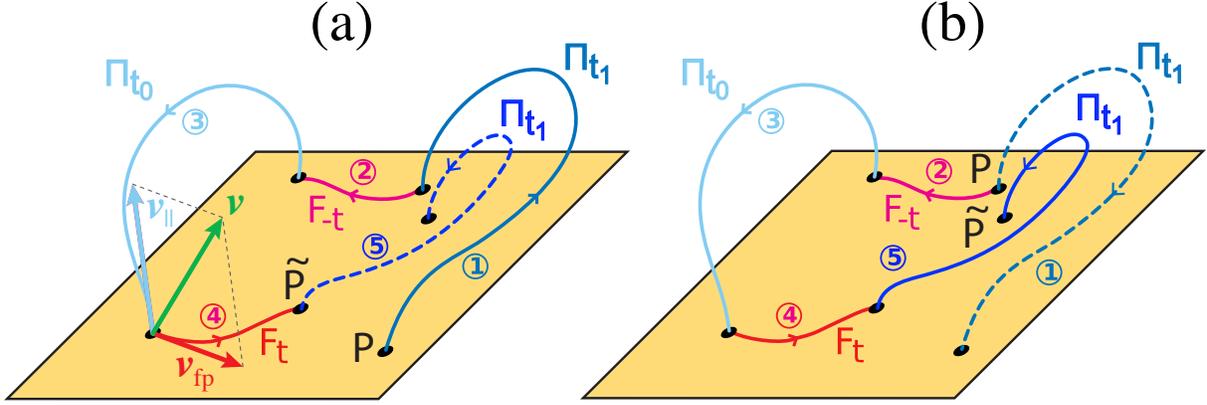}
\caption
{A pointwise definition of the extended slip-back mapping (a) and its dual analog (b) in a closed-field region;
the slip-back mapping itself is composed of
four distinct mappings that connect a given point $\mathrm{P}$
to its image point $\mathrm{\tilde{P}}$, both referred to a current time $t_1$.
In case (a), the connectivity change due to this composite mapping becomes fully determined by the field line (dashed line) traced from the image point $\tP$.
Altogether, this is accomplished in the following five steps:
1)  At $t=t_1$, trace the field line from a given footpoint $\mathrm{P}$ at the boundary;
2) Advect its conjugate footpoint backward in time from $t_1$ to $t_0$;
3)  At time $t_0$, trace the field line from the advected point;
4) Advect its conjugate footpoint along the boundary forward in
time from $t_0$ to $t_1$ to reach a point $\mathrm{\tilde{P}}$, which defines
the result of the slip-back mapping;
5) Reveal its connectivity significance by tracing the field line from $\mathrm{\tilde{P}}$ at $t=t_1$.
Both footpoint advections are made here through the ideal MHD footpoint velocity field $\vfp$ (see Eq.~\ref{vfp}). 
In the second scheme (b), the slip-back mapping starts with the advection backward in time applied to a given point $\mathrm{P}$, which appears in the scheme (a) as a conjugate footpoint of the first field line at $t=t_1$ (dashed line in panel (b)).
The remaining three steps of this mapping are the same as in the five-step mapping (a), so that $\mathrm{\tilde{P}}$ coincides now with the point that appears in the scheme (a) as a conjugate footpoint of the second field line at $t=t_1$ (dashed blue line in panel (a)).
\label{fig-5steps}
}
\end{figure*}

At each boundary sphere, we define a $N_\theta \times N_\phi$ footpoint mesh, $\theta_j,\phi_k ; j=1,N_\theta ; k=1,N_\phi$, which may be nonuniform and not necessarily the same for both spheres.
We create on these meshes for the current time $t_1$ color-coded maps that classify the corresponding footpoints of the field lines by the type of reconnection they undergo within a given time interval $[t_0, t_1]$, where $t_0$ is the initial time of the field evolution.
To identify the type of reconnection {\hB during} this evolution, we map each mesh point $\pP$
% in five steps 
by following one of the two schemes shown in panels (a) and (b) of Figure~\ref{fig-5steps}
for the simplest case of a closed-field region
that does not contain footpoints approaching or emerging at the PIL during the time period under consideration.
The scheme shown in panel (a) implies taking the following five steps:
\begin{enumerate}

\item Using $\mathbf{B}(r,\theta,\phi,t_1)$, one traces the field line (cerulean {\hB color} in Figure~\ref{fig-5steps}(a)) from a given mesh point $\pP$, which maps $\pP$ to its {counterpart} or conjugate footpoint {and} thereby defines the corresponding field-line mapping $\Pi_{t_1}$.

\item The obtained conjugate footpoint, $\Pi_{t_1}(\pP)$, is then advected along the boundary backward in time to 
{the initial}
instant $t_0$ via the {\hB negative of the} footpoint velocity field 
defined by Equation (\ref{vfpa}).
The mapping resulting from this advection is denoted as $F_{-t}$ in Figure~\ref{fig-5steps}(a).

\item Then,  using $\mathbf{B}(r,\theta,\phi,t_0)$, one
{launches}
the field line from point $F_{-t} \circ \Pi_{t_1}(\pP)$, which maps
{the latter}
to the corresponding conjugate footpoint $ \Pi_{t_0}  \circ  F_{-t} \circ \Pi_{t_1}(\pP)$.
This field line is colored in sky blue in Figure~\ref{fig-5steps}(a) and the corresponding field-line mapping is denoted as
$\Pi_{t_0}$.

\item The newly obtained conjugate footpoint, $ \Pi_{t_0}  \circ  F_{-t} \circ \Pi_{t_1}(\pP)$, is then advected along the boundary forward in time from $t_0$ to $t_1$ {with} the same footpoint velocity as before.
Thus, one obtains a point $\tP=F_{t}\circ \Pi_{t_0}  \circ  F_{-t} \circ \Pi_{t_1}(\pP)$ that is the image of the mesh point $\pP$ due to
{the}
slip-back mapping as originally defined by \citet{2009ApJ...693.1029T}.

\item To analyze reconnection in such a simple case as this, the four-step composition of mappings we described
{would already be}
 sufficient.
However,  one {\hB  more step} is required to make possible the analysis of more general configurations containing not only closed but also open and disconnected field lines.
Namely, by using $\mathbf{B}(r,\theta,\phi,t_1)$ one 
{\hB launches} 
a field line from $\tP$ to some $\Pi_{t_1}(\tP)$ to fully identify the significance of the connectivity change in the composed four-step mapping, which will become clearer later on.
The corresponding field line is shown dashed blue in Figure~\ref{fig-5steps}(a); it defines the fifth mapping in the resulting composition.

\end{enumerate}
%=======================================================
%=======================================================
%
Thus the first four steps pointwise define here the original slip-back mapping of \citet{2009ApJ...693.1029T} 
\begin{equation}
S_{-t}= F_{t}\circ \Pi_{t_0}  \circ  F_{-t} \circ \Pi_{t_1} \, ,
	\label{Smt}
\end{equation}
while composing it further with $\Pi_{t_1}$, one defines a composite mapping
\begin{equation}
\tilde{\Pi}_{t_1}= \Pi_{t_1} \circ S_{-t} \, ,
	\label{tPt1}
\end{equation}
which we henceforth call the {\it extended slip-back mapping} or briefly the five-step mapping.

Figure~\ref{fig-5steps}(b) illustrates a pointwise definition of another mapping, which we call {\it dual slip-back mapping}, defined as
\begin{equation}
S^{*}_{-t}=  \Pi_{t_1} \circ F_{t}\circ \Pi_{t_0}  \circ  F_{-t} \, ,
	\label{Smtd}
\end{equation}
where the advection backward in time is applied to the mesh point $\pP$ itself rather than to its image $\Pi_{t_1}(\pP)$,
as it takes place in the $S_{-t}$ mapping.

The remaining three steps in this dual mapping $S^{*}_{-t}$ are the same as in the five-step mapping 
presented in Figure~\ref{fig-5steps}(a)
and defined by Equations (\ref{Smt}) and (\ref{tPt1}).

The connectivity change due to the  $S^{*}_{-t}$ mapping, involving generally closed, open, and disconnected field lines, becomes fully determined 
{after}
 verifying the connectivity of the field line (dashed cerulean line in panel (b)) {launched} from point $\pP$ rather than $\tP$.

In ideal MHD, {as a consequence of the frozen-in law condition}, 
every point $\pP$  must coincide with its image $\tP$, irrespective of  which of the two mappings, $S_{-t}$ or $S^{*}_{-t}$, has been applied.
{In other words}, the corresponding extended slip-back mapping and the field-line mapping at the time $t_1$ must coincide with each other,
\begin{eqnarray}
\tilde{\Pi}_{t_1} & \equiv & \Pi_{t_1}\, ,
\end{eqnarray}
while $S_{-t}$ and $S^{*}_{-t}$ must be both the identity mapping,
\begin{eqnarray}
S_{-t} = S^{*}_{-t} \equiv \mathrm{Id} \, .
\end{eqnarray}
However, for a real plasma with a nonzero resistivity, 
these relations may hold only approximately or even completely break down in certain regions of the magnetic configuration.
Then the incurred deviations of $\tilde{\Pi}_{t_1}$ from $\Pi_{t_1}$, or {deviations} of $S_{-t}$ and $S^{*}_{-t}$ from the identity mapping $\mathrm{Id}$, can be used to identify and even to quantify the 
 reconnection processes in those regions.

Indeed, the definition of the five-step mapping $\tilde{\Pi}_{t_1}$ implies that 
{every mesh point $\pP$ is associated with}
a sequence of three field lines that represents a change of {their} connectivity 
{during a given time interval as a result of resistive slippage of Lagrangian plasma elements}.
In general, these field lines may appear in the sequence as closed (\textbf{C}), open (\textbf{O}), or even disconnected (\textbf{D}), depending {on the boundary to which $\pP$ belongs} and the type of underlying {magnetic reconnection or diffusion} process.
The same conclusion equally refers to the five-step mapping based on the dual slip-back mapping $S^{*}_{-t}$ (Figure~\ref{fig-5steps}(b)).
As will become clear below, this mapping allows us to identify
the counterparts of
the footprints of reconnected flux tubes determined by $\tilde{\Pi}_{t_1}$.
We will see in Section \ref{sec-application} that
such a complementarity between the mappings is important, especially in the cases when the calculation of one of these mappings is problematic.

%, except that  in this case one needs to swap the closed and disconnected field lines.
\subsection{Classification of Connectivity Changes}
\label{ssec-class}

Consider how one classifies the connectivity changes in more detail under the simplifying assumption that the footpoints advected by the $F_{-t}$ and $F_{t}$ flows do not approach or start at the PIL during the time interval of interest.
Simple combinatorics shows that
%yields for each of the two boundaries only eight different combinations of the field lines involved in the described $\tilde{\Pi}_{t_1}$ mapping.
%In other words, 
the connectivity of magnetic field lines at a given boundary can be changed {in this case}
% in a non-ideal magnetic evolution 
only in eight distinct ways,
which are uniquely identified by three-letter code words composed of the above
 \textbf{O}, \textbf{C}, and \textbf{D}.
% by representing the above-mentioned sequence of the field lines.
The first and third letters in the code words represent the types of the field lines with the footpoints at $\pP$ and $\tP$, respectively,  at time $t_1$.
The second letter represents the type of the field line at time $t_{0}$.
Therefore, the code word is to be read from left to right in the same order as the field lines {consecutively} appear in the five-step mapping.

%admits only eight possible time variations of the field-line connectivity to a given boundary surface.
\contourlength{0.3pt}
\contournumber{3}

\begin{figure*}[t]
\centering
\includegraphics[width=0.9\textwidth]{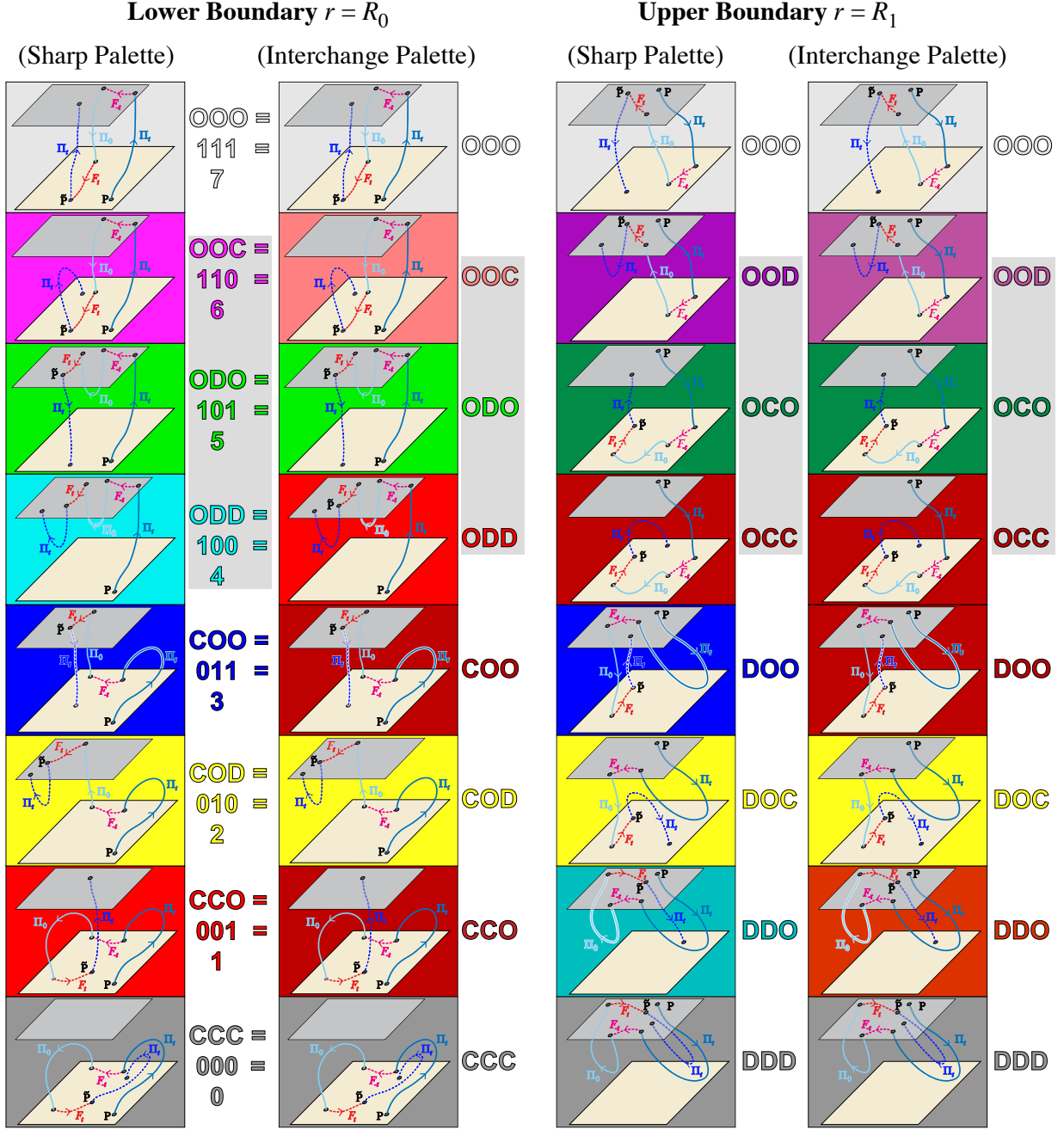}
\caption
{All possible changes of the field-line connectivity that can occur within
a given time period for a point $\pP$ at the lower and upper radial boundaries
%, $r=R_0$ and $r=R_1$,
during a resistive evolution of the plasma-magnetic configuration.
The letters and bits (0 for \textbf{C} or \textbf{D}, and 1 for \textbf{O}) represent the type of the field lines associated with $\pP$ in the extended slip-back mapping, as depicted in Figure~\ref{fig-5steps}(a) and explained in Sections~\ref{ssec-concept} and \ref{ssec-class}.
The corresponding areas of the connectivity changes are depicted on the slip-back maps by using either sharp or interchange palettes, different for each of the boundaries; the palettes are shown here via background colors of the plates representing the cases.
The gray-shaded code words designate the cases that can appear at the counterpart boundaries in the maps based on the dual slip-back mapping (Figure~\ref{fig-5steps}(b)), which generally raises the number of possible cases at each boundary from eight to eleven.
}
\label{fig-classification}
\end{figure*}

To visualize these changes of the connectivity, we associate every mesh point $\pP$ at the boundary with the corresponding code word and a certain color.
Figure~\ref{fig-classification} provides the full classification of the changes for both boundaries by illustrating each class with a diagram similar to the one shown in Figure \ref{fig-5steps}.
The background of these diagrams is depicted in the associated color of two different sets, which we call sharp and interchange palettes.

The sharp palette is designed to facilitate discrimination between neighboring mesh points of different classes {\hB by employing high-contrast colors.}
Using this palette,  we obtain for the lower boundary $r=R_0$ the following classification: \CCC, \CCO, \COD, \COO, \ODD, \ODO, \OOC, and \OOO\ (first column in Figure \ref{fig-classification}).
Analogous classification for the upper boundary $r=R_1$ yields \DDD, \DDO, \DOC, \DOO, \OCC, \OCO, \OOD, and \OOO\ (third column in Figure \ref{fig-classification}), where a similar sharp palette is applied.
%The \CCC\ and \DDD\ classes are chosen to be colored the same in both classifications; the same is true for the \OOO\ class,
%{\hR
% while to the other classes we applied a similar sharp palette with lighter colors.
%}
The colors are distributed over the classes so as to indicate the complementarity of the corresponding changes of the field-line connectivity to the boundaries,
which facilitates the interpretation of the resulting maps.

%=======================================================
%=======================================================

In this respect, we found {\hB it} helpful to use an alternative color code based on a so-called interchange palette.
{As the name hints,
this palette} is designed to emphasize the connectivity changes {caused by} interchange reconnection.
The second column in Figure~\ref{fig-classification} presents the corresponding {lower-boundary} classification, {whose code words are colored as follows:} \CCCi, \CCOi, \CODi, \COOi, \ODDi, \ODOi, \OOCi, and \OOOi.
The interchange palette for the upper boundary provides \DDDi, \DDOi, \DOCi, \DOOi, \OCCi, \OCOi, \OODi, and \OOO\ (fourth column in Figure \ref{fig-classification}).

The red hue colors in these palettes designate the connectivity changes produced {exclusively by interchange reconnection, which develops between either open and closed {\huge}field lines or  open and disconnected field lines}.
However, it should be noted that such field lines are only conditionally open or disconnected, because the topological classification of the field lines at the helmet-streamer core may change {depending on how far from the Sun one sets the upper boundary}.

With this reservation, the connectivity changes in our classification are definitely topological {if their code words contain at least two different letters}.
Moreover, almost all of these topological changes are caused by magnetic reconnection.
The exception is the \OCO\ class of changes, which are  {produced mainly} by the solar-wind advection of {originally} closed field lines through the upper boundary.
Needless to say, the opening of closed field lines {at the core of the helmet streamer} is only conditional {in this process}, since field lines that are identified as open in one coronal model can still be closed in another model with identical parameters but larger radius $R_{1}$.

In contrast, the opening becomes truly topological when a closed field line {in the helmet streamer} approaches a magnetic null point and reconnects there with a disconnected field line {by producing} two open field lines.
That having been specified, we will no longer {return to the discussion on} which connectivity changes are conditionally or truly topological.
The topological changes of the {described} type belong to the \ODO\ class and {are revealed} by the slip-back maps at the lower boundary.

In general, the reconnection of this kind
 is more likely to occur {along} separator field lines that {lie at the intersection of separatrix surfaces that delimit \textbf{O}, \textbf{C} or \textbf{D} flux regions.
If the radius of the upper boundary $R_1$ is set to be large enough, these surfaces are fan separatrix surfaces of the null points that likely spread all over the edge of the disconnected flux regions in the helmet streamer.
At smaller values of $R_1$, some of the fans turn into so-called ``bald-patch'' separatrix surfaces, which consist of the field lines touching the upper boundary
\citep{2017ApJ...845..141T}.
In this case,
the intersection of the separatrix surfaces also provides a separator field line along which the reconnection involving the \textbf{O}, \textbf{C}, and \textbf{D} types of field lines can take place.

Knowing \OCO\ areas at the upper boundary and \ODO\ areas at the lower boundary
{\hB allows us to} estimate the magnetic flux that becomes open for a given time interval $[t_{0},t_1]$ purely due to the solar-wind advection.
This flux is simply given by the difference between the fluxes calculated for the \OCO\ and \ODO\ areas}.

For \CCC\ and \OOO\ areas, {no connectivity changes can be inferred without a more sophisticated analysis}. 
%%
%{\hR
%\begin{verbatim}
%Not entirely true!!!
%Only if P and tilde P are "sufficiently close" to each other.
%A large gap between them implies that either 1) a chain of reconnections
%occurred within a given time interval but missed because of large t-t0
%or 2) a reconnection that did not lead to the change of connectivity
%between the boundaries.
%\end{verbatim}
%}
In the case of \CCC, both footpoints of a presently 
closed field line appear to be associated with previously closed field lines.
Depending on the length of the time interval, $t_1 - t_0$,
it is possible {in principle} that two mutually canceling topological changes might have happened (e.g., opening up {first and then} closing down of the field line).
The ``trivial'' \CCC\  region might also contain a null point or QSL that is passed through by some plasma elements connected {via a field line} at $t=t_1$ to $\pP$ and thereby {participated} in {a} reconnection of closed field lines.
{Analogously in case \OOO, no changes of the connectivity type occur in the corresponding slip-back mapping.
If point $\pP$ in the \OOO\ or \CCC\ area is separated from its image $\tP$ by a distance {that varies smoothly as a function of $\pP$}, it is 
likely evidence of magnetic diffusion rather than reconnection
(see {\hB Subsection \ref{subsec-slippages} 
for a discussion on the nature of the error}).
Conversely, a jump in the distribution of the distances would imply a reconnection in the corresponding location. 
Thus, the analysis of the slippage distances or even more elaborate measures such as slip-squashing factors would allow one to identify reconnection sites in the \OOO\ and \CCC\ cases \citep{2009ApJ...693.1029T}.
{\hB However, such in-depth analysis is beyond the scope of the present study,
in which we would like to detect magnetic reconnection 
that leads to change of connectivity type only.}

In contrast with  the above-considered case \ODO, which describes the opening of closed field lines, case \COD\ represents the closing down of open field lines.
{\hB One such open field line} is shown at time $t_0$ in Figure~\ref{fig-classification} as a sky blue line.
Apparently, it has a counterpart open field line at this time with which it reconnects later at some $t\leq t_1$.
The result of the reconnection is one closed and one disconnected field line shown in Figure~\ref{fig-classification} at time $t_1$: the closed (cerulean) and disconnected (dashed blue) field lines are rooted at points $\pP$ and $\tP$, respectively.

Cases \CCO, \COO, \OOC, and \ODD\ represent various types of interchange reconnection.
{For example,} case \CCO\ {includes} at time $t_1$ one closed (cerulean) and one open (dashed blue) field line.
At time $t_0$, they originate from a closed field line (sky blue) whose one footpoint later appears to be a footpoint of the closed field line while the other footpoint becomes the lower-boundary footpoint of the open field line.
{From this} we deduce that interchange reconnection has {likely} occurred during the time interval $[t_0, t_1]$ between the (sky-blue) closed field line and some open field line {(not shown in the \CCO\ diagram in Figure~\ref{fig-classification})}.

{Similarly to case \CCO, case \COO\ includes at time $t_1$ one closed (cerulean) and one open (dashed blue) field line.  However, at time $t_0$, they originate from an open (sky blue) rather than a closed field line.
Its lower-boundary footpoint later appears to be a footpoint of the closed field line while its upper-boundary footpoint becomes a footpoint of the the open field line.
Thus, interchange reconnection has likely occurred during the time interval $[t_0, t_1]$ between the (sky-blue) open field line and some closed field line {(not shown in the \COO\ diagram in Figure~\ref{fig-classification})}.

 Cases \OOC\ and \COO\ are very similar: they both include at time $t_1$ open and closed field lines, which originate at time $t_0$ from an open field line (sky-blue).
One case is obtained from the other by reversing the order in which the field lines appear in the corresponding slip-back mapping.
In both cases we have the same change of connectivity, which implies that the interchange reconnection has likely occurred during the time interval $[t_0, t_1]$ between the open field line and some closed field line (not shown on the \OOC\ diagram in Figure~\ref{fig-classification}). 

Case \ODD\ includes at time $t_1$ open and disconnected field lines, whose one of the two footpoints are $\pP$ and $\tP$, respectively.
At time $t_0$, these field lines originate from a disconnected field line (sky blue) whose one footpoint later appears to be the upper-boundary footpoint of the open field line, while the other footpoint becomes a footpoint of the disconnected field line.
Thus, interchange reconnection has likely occurred during the time interval $[t_0, t_1]$ between the (sky-blue) disconnected field line and some open field line  (not shown in the \ODD\ diagram in Figure~\ref{fig-classification}).

This completes the classification of the connectivity changes that evolving magnetic configurations experience sufficiently far from the PIL at the lower boundary.
Simple substitution of \textbf{C} for \textbf{D} in this classification provides a similar classification of the connectivity changes for the upper boundary (see the two rightmost columns in Figure~\ref{fig-classification}).

As already discussed for the \CCC\ and \OOO\  cases, one has to remember that all classified connectivity changes are, by definition, only effective.
For their simplest interpretation, we can employ a single-reconnection scenario.
However, by using just slip-back maps we may not generally exclude that multiple reconnection events occurred during a given time interval with effectively the same result of a single event.
To resolve this uncertainty, one needs to apply extra field-line mapping techniques as mentioned above.

\subsection{Generalization of the Slip Mapping Technique}

The described technique of tracking field lines {by means of} the footpoint velocity $\vfp$ {that is} defined and approximated by Equations (\ref{vfp}) and (\ref{vfpa}), respectively, has one {restriction}: it allows {us} to identify the connectivity changes at the boundary surfaces only in the regions sufficiently far from the PILs.
In spite of this restriction, the technique still happens to be very useful to understand idealized MHD evolutions whose driving velocity at the lower boundary has only a tangential component (Section \ref{sec-application}).

However, to track field lines in more realistic evolutions with emerging or submerging magnetic fluxes at all boundaries, {this} technique must be generalized.
This may be done by modifying $\vfp$ and extending it to the volume to obtain what we call
the tracking-field-line velocity field,
\begin{eqnarray}
  \vtfl =  \mathbf{v} 
        - \frac{v_{r}\, B_{r}\, \mathbf{B}}{{B_{r}^2}+\epsilon^2}
  \, .
	\label{vtfl}
\end{eqnarray}
We propose to use $\vtfl$ to track Lagrangian particles backward and forward in time, which will provide us with the required tracking of the field lines in the slip-back mapping.

Indeed, the subtraction of any field-aligned component from the full velocity does not violate the frozen-in law condition.
In the flows $F_{-t}$ and $F_{t}$ calculated with such a velocity, the Lagrangian particles would continually slide from one plasma element to another by remaining nevertheless on the same field line {in an} ideal MHD evolution.
The particular choice of the field-aligned component in Equation (\ref{vtfl}) ensures that
the tangential component of $\vtfl$ coincides at each boundary with our approximate $\vfp$ given by Equation (\ref{vfpa}).
The radial component of $\vtfl$ is
\begin{eqnarray}
  \hat{\mathbf{r}} \cdot \vtfl = 
        \frac{v_{r}\, \epsilon^2}{{B_{r}^2}+\epsilon^2}
    \, ,
	\label{vrtfl}
\end{eqnarray}
which decreases at $B_{r} \gg \varepsilon$ quadratically with $\varepsilon$ and has maximum value $v_{r}$ at vanishing $B_{r}$.
Thus, choosing $\vtfl$ for Lagrangian particles that were originally located at the mesh points of the boundaries,  one can keep them close to the boundaries everywhere {except for} a narrow layer around the surface $B_{r}=0$ controlled by {a small} value of $\varepsilon$. 
By also using these particles as launch points to trace field lines at the {instances} $t_0$ and $t_1$, one {will} reveal the corresponding field-line connectivity to the boundaries at these {instances} and hence its change in our slip-back mapping.

The same velocity field $\vtfl$ can be used without a modification to track field lines in the slip-forth mapping, which was constructed to identify at $t=t_0$ the flux tubes that are going to reconnect during a given time interval $[t_0, t_1]$ \citep{2009ApJ...693.1029T}.
The {respective} classification of the connectivity changes {can be obtained from that} shown in Figure \ref{fig-classification} {by simply swapping} the flows $F_{-t}$ and $F_{t}$ as well as the {instances} $t_0$ and $t_1$ {in the corresponding diagrams}.

The implementation of these generalized slip mappings based on the use of $\vtfl$ will be described in a future paper.
Here we assess
in the  potential of this technique by using its simpler version in which field lines {are tracked} through their footpoints.

\section{AN APPLICATION OF SLIP-BACK MAPPING}
\label{sec-application}
%%%%%%%%%%%%%%%%%%%%%%
%%%%%%%%%%%%%%%%%%%%%%
%
We  apply the slip-back mapping to an idealized simulation
of differential rotation, following \citet{2005ApJ...625..463L} and 
\citet{1996Sci...271..464W}. \citet{2005ApJ...625..463L}  
performed a 3D MHD simulation of differential rotation on a
$101\times 101 \times 181$ ($r,\theta,\phi$) mesh using the
polytropic model.
{This simulation was made in order to understand how} the reconnection that occurs at {the boundaries of coronal holes allows them to keep} their almost rigid rotation, as proposed by \citet{2004ApJ...612.1196W}.
The differential rotation profile is similar to that used in 
\citet{1996Sci...271..464W}, but the differential rotation part is
sped up by a factor of ten to reduce the computation time:
\begin{equation}
\omega(\theta) = 13.39 -2.77 \cos^2 \theta \deg \mathrm{day}^{-1},
\label{eq-diff_rot}
\end{equation}
i.e., the multiplier 2.77 is increased to 27.7.
%=======================================================
\begin{figure}[t]
\centering
\includegraphics[width=0.95\textwidth]{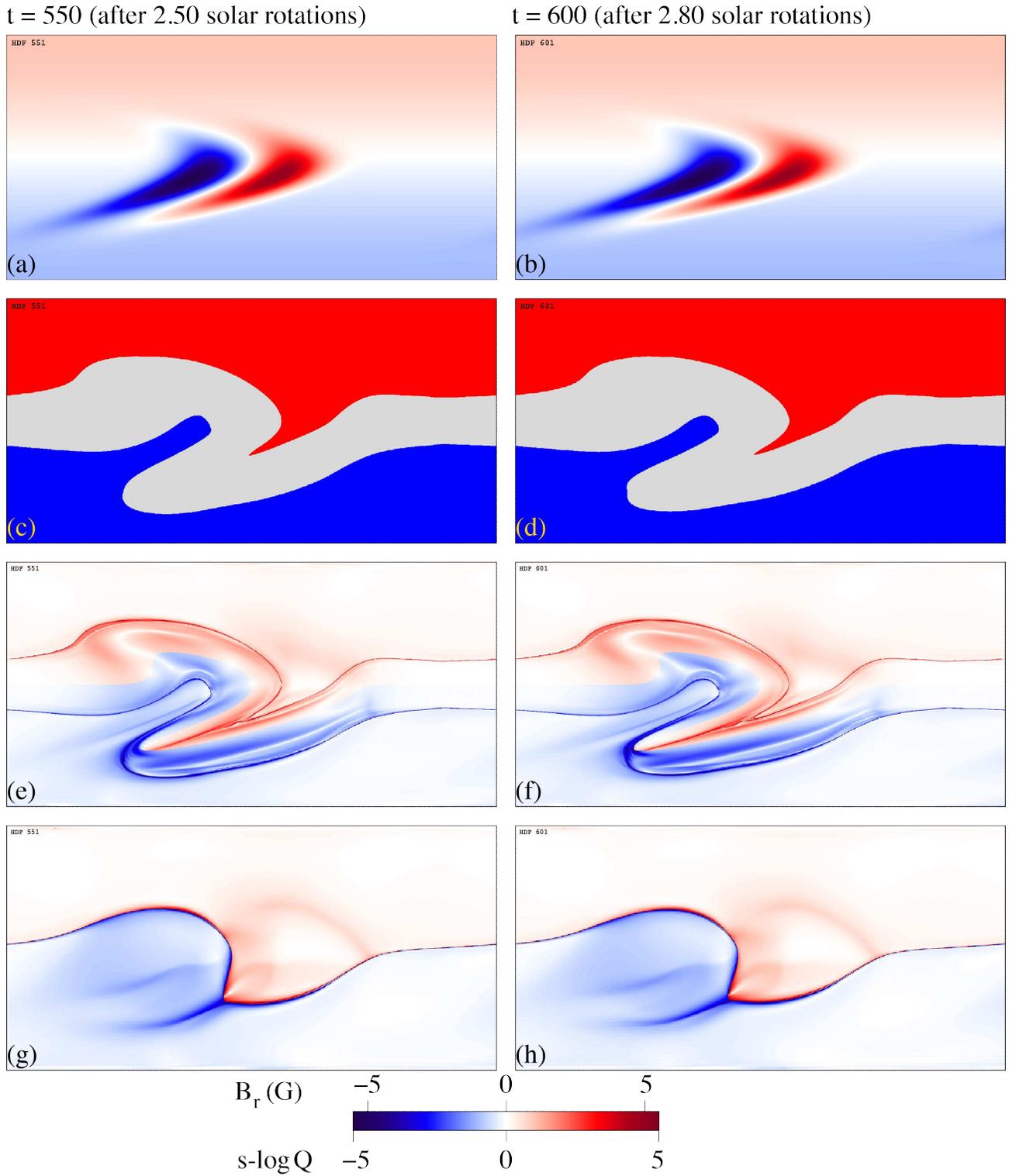}
\caption
{Top: the magnetic field at the solar surface (red positive, blue negative) at $t=550$ {\hB (a)} and $t=600$ {\hB (b)} code units (corresponding
to 2.5 and 2.8 solar rotations) during
the simulation of \citet{2005ApJ...625..463L}. Middle: 
coronal hole maps (red: open positive field; blue: open negative field;
gray: closed field; {\hB at $ t=550$ (c)} and $t=600$ {\hB (b))}  Bottom: maps of $\slog Q$, which is the so-called signed logarithm of the squashing factor $Q$; its sign coincides with the sign of the local $B_{r}$; 
 {\hB (e) $ t=550$ at $R_\sun$,  (f) $t=600$ at $R_\sun$, 
(h) $t=550$ at $19~R_\sun$, (g)  $t=600$ at $19~R_\sun$. }
}
\label{fig-br_ch_q}
\end{figure}
%=======================================================

%%%%%%%%%%%%%%%%%%%%%%
%%%%%%%%%%%%%%%%%%%%%%
%
Figure~\ref{fig-br_ch_q} shows the radial magnetic field, the coronal hole map, 
and the squashing factor \citep{2007ApJ...660..863T} at the solar surface for
two different times during the simulation, 550 code units (corresponding
to 2.50 solar rotations) and 600 c.u.\ (corresponding to 2.80 {\hB solar rotations}).
 The bipolar $B_r$ distribution 
is progressively elongated from the original circular shapes. The coronal
hole map suffers deformation to a lesser extent \citep{2004ApJ...612.1196W}.
The squashing factor maps {reveal} the footpoints of field lines associated
with separatrices and quasi-separatrices, which are generally favorable sites for the development of magnetic reconnection \citep{2007ApJ...660..863T}.
They also include the boundaries between open and closed field regions.
{\hB We now proceed to calculate the slip-back maps relative to the
magnetic evolution portrayed in Fig.~\ref{fig-br_ch_q}, evaluate
footpoint slippages that are present in the evolution, and analyze the
topological structure of the field lines associated with a particular
class region.}
%
%=======================================================
\begin{figure*}[t]
\centering
\includegraphics[width=0.9\textwidth]{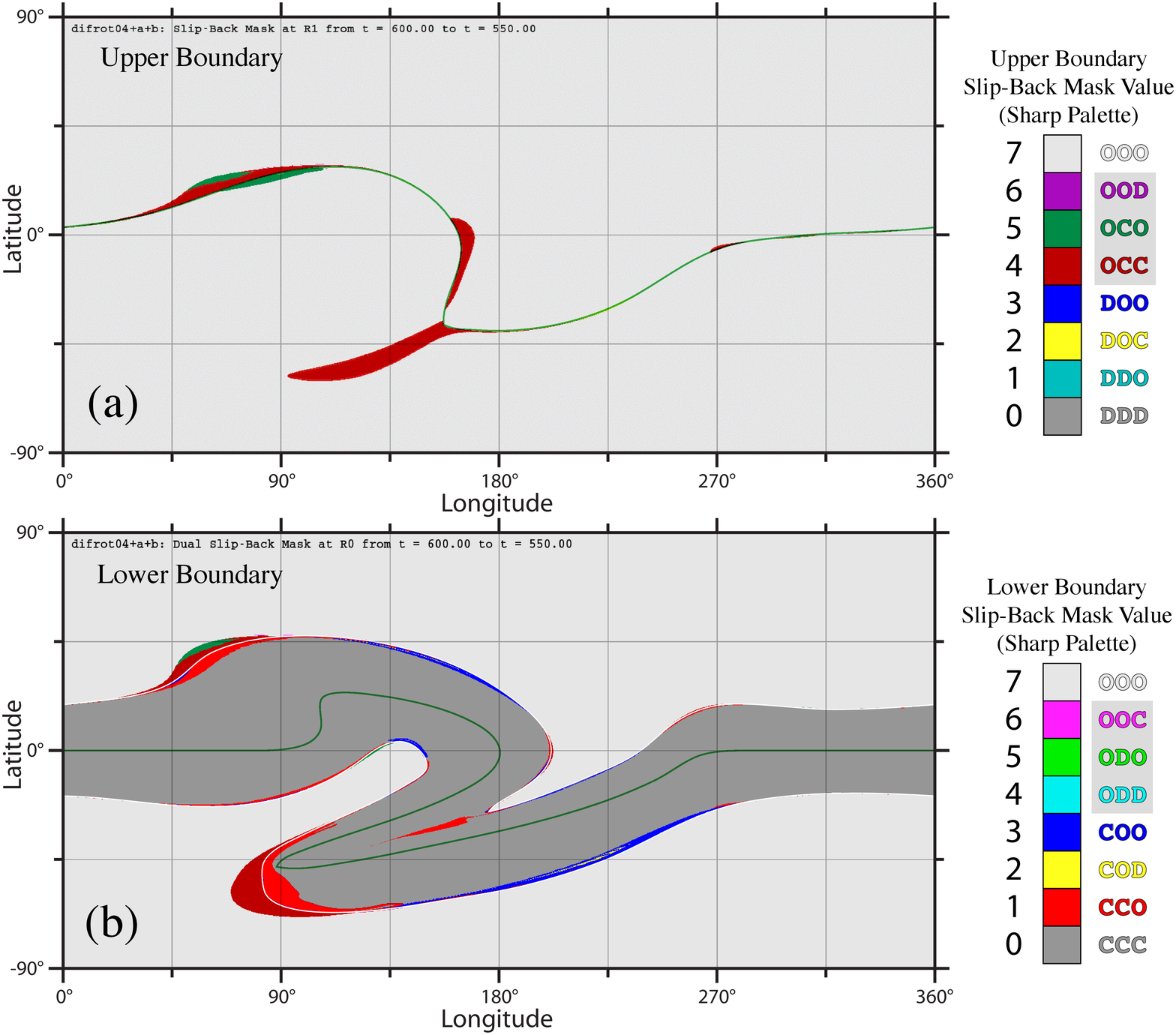}
\caption
{Slip-back maps for the simulation of
\citet{2005ApJ...625..463L} calculated between
550 c.u.\ (2.50 solar rotations) and 600 c.u.\ (2.80 solar rotations) at
the upper (a), {\hB $19~R_\sun$}, and lower (b) boundaries.
Both maps use the sharp palettes that clearly discriminate various cases of
the connectivity changes shown in Figure~\ref{fig-classification}. 
The connectivity changes whose code words are shaded in gray may appear
at the counterpart boundary when the dual slip-back mapping is applied.
The color palettes of the maps are logically complementary to
each other; darker hues are used for the \OOD, \OCO, \OCC, and \DDO\ regions
to distinguish them from the \OOC, \ODO, \CCO\ and \ODD\ regions when the region
and its counterpart are both present on the same map.
The green lines at both boundaries show PILs, the white lines at the lower boundary depict the coronal hole boundaries.
}
\label{R0_R1_SB_maps}
\end{figure*}
%%%%%%%%%%%%%%%%%%%%%%
%=======================================================
\begin{figure*}[t]
\centering
\includegraphics[width=0.9\textwidth]{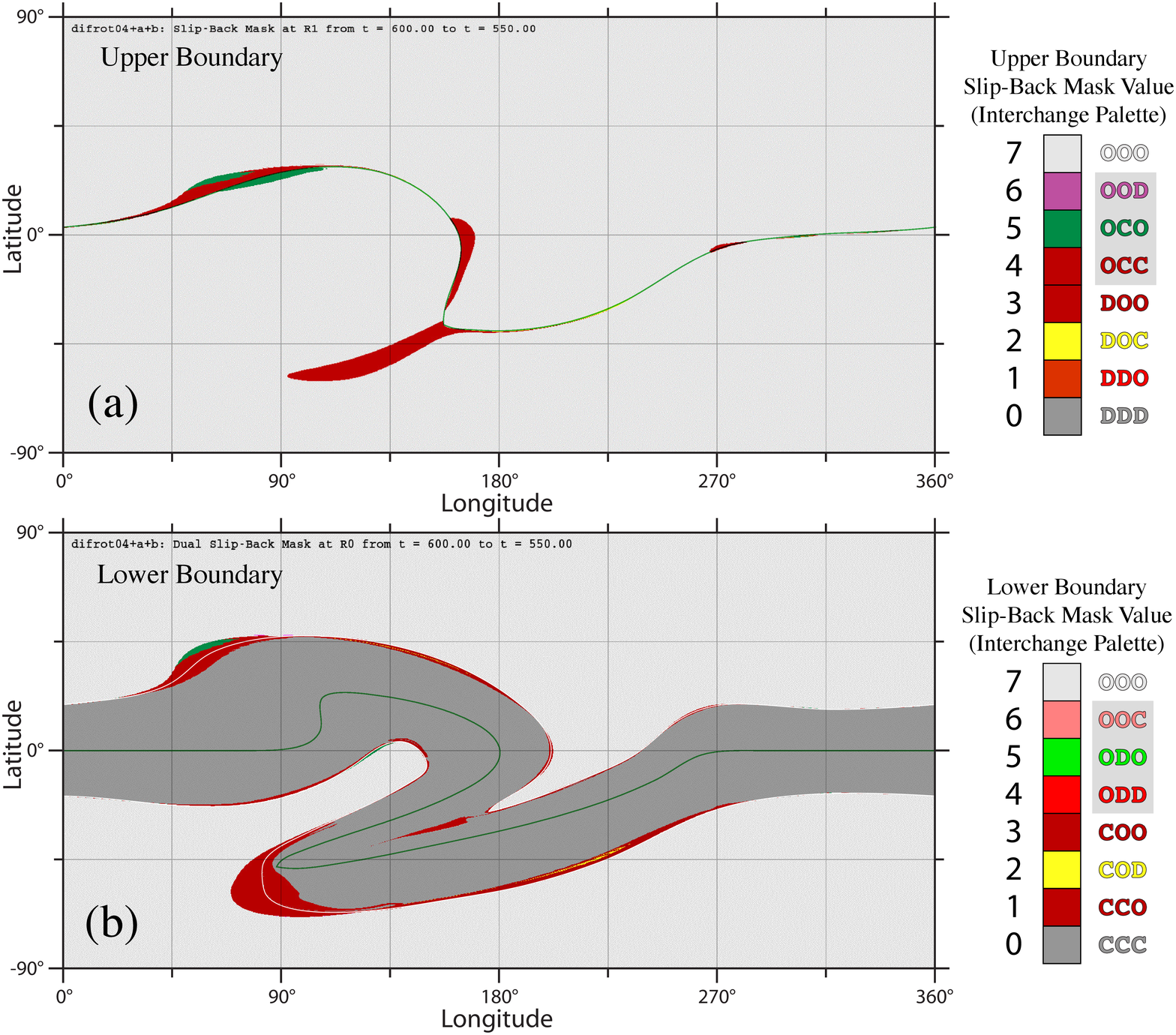}
\caption
{Slip-back maps for the simulation of
\citet{2005ApJ...625..463L} calculated between
550 c.u.\ (2.50 solar rotations) and 600 c.u.\ (2.80 solar rotations);  the interchange palettes are used (Figure~\ref{fig-classification}), which group together regions of interchange reconnection (\OODi, \OCCi, \DOOi, \DDOi,
\OOCi,  \ODDi,  \COOi, and \CCOi)
with various red hues.
The connectivity changes whose code words are shaded in gray may appear
at the counterpart boundary when the dual slip-back mapping is applied.
%The color palettes of the maps are logically complementary to each other; darker hues are used for the \OODi, \OCOi, \OCCi, and \DDOi\ regions to discriminate them from the \OOCi, \ODOi, \CCOi\ and \ODDi\ regions when the region and its counterpart are both present on the same map.
The green lines at both boundaries show PILs; the white lines at the lower boundary depict the coronal hole boundaries.
{\hB The upper boundary is at  $19~R_\sun$}.
}
\label{R0_R1_SB_i_maps}
\end{figure*}
%%%%%%%%%%%%%%%%%%%%%%
%=======================================================
\begin{figure*}[t]
\centering
\includegraphics[width=0.7\textwidth]{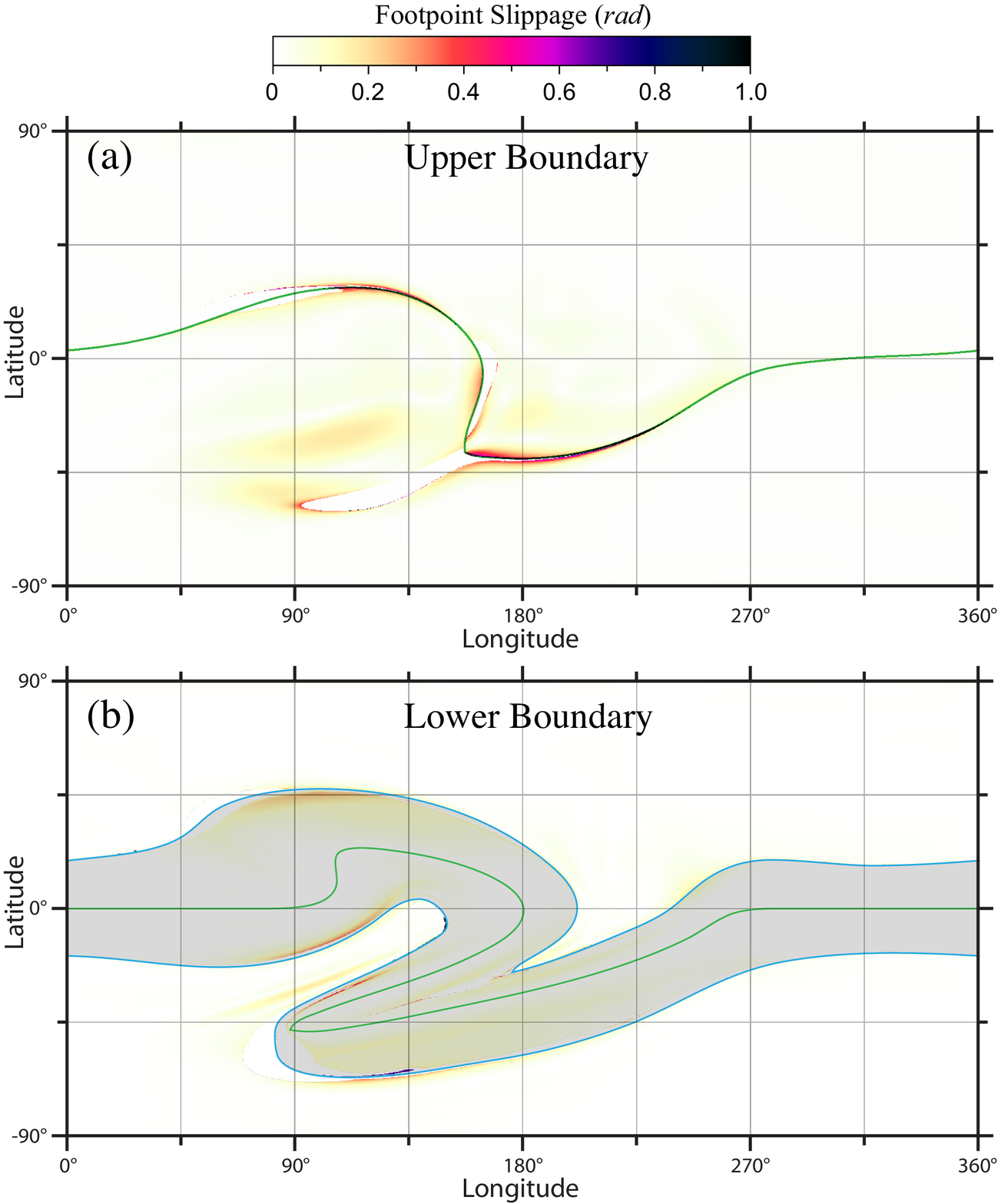}
\caption
{Footpoint-slippage angular distance in the \OOO\ regions at the upper boundary {\hB at $19~R_\sun$,} (a) and in the \CCC --\OOO\ regions at the lower boundary (b). The light-grey-shaded area corresponds to the closed field regions; the cyan lines at the lower boundary show the coronal hole boundaries, the PILs are colored at both boundaries in green.
}
\label{fig-fp_slip}
\end{figure*}
%%%%%%%%%%%%%%%%%%%%%%
\subsection{\hB Calculation of the Slip-Back Maps}
For the time interval shown in Figure~\ref{fig-br_ch_q},
we calculate the slip-back maps at the boundaries at $r=R_0\equiv R_\sun$ and $r=R_1\equiv 19~R_\sun$.
{\hB The value of  $19~R_\sun$ is chosen to be less than the actual boundary radius at $20R_\sun$  to avoid possible boundary
effects. }
The results we obtained are presented by using sharp and interchange palettes (see Figures~\ref{R0_R1_SB_maps} and \ref{R0_R1_SB_i_maps}, respectively).
These maps are produced on the mesh whose points {are uniformly distributed over} spherical boundaries {with an angular separation of} $0.25\arcdeg$, both in latitude and in longitude.

To produce the upper-boundary maps, we applied only the extended slip-back mapping and not its dual analog.
Using the latter would imply tracking field lines via {advection of their} footpoints backward in time, which means {the closing down of} field lines all over the PIL  (except for disconnected field regions {where an opposite process likely takes place}).
The field-line tracking then becomes impossible after converging footpoints 
reach the PIL.
Therefore, our upper-boundary maps do not contain regions of the \OOC\ type.
Fortunately, the prominent connectivity changes of the \OCC\ and \OCO\ types are well recovered at the upper boundary, because the backward-in-time advection of the footpoints occurs in these cases at the lower boundary, where the footpoint flow is well defined during the entire time interval $[t_0, t_1]$.

The regions of interchange
reconnection (\CCO, \COO, \ODD, \OCC, and  \OOD) 
represent together less than
 2\% of the total area (although no \ODD\ points are found to be present)
and are concentrated  at the
boundaries between the persistently open (\OOO, 54\% of the total area) and
persistently closed field
regions (\CCC, 42\%).
The
\COD\ regions (closing down of two open field lines) and
 \OCO\ (opening up of closed field lines mainly via solar-wind advection) account for only 0.01\%
of the area.

The color palettes used for the upper-boundary maps are logically complementary to
those used for the lower-boundary map.
In fact, \DDO\ and \OCC\  at
$r=19~R_\sun$ respectively correspond
 to \ODD\ and \CCO\ at $r=R_\sun$.
Unsurprisingly, 98\% of the area
 at the upper surface is classified \OOO\ (i.e., lines
{\hB were and remain} open). \OCC, \OCO, \DOC, \OOD,
and \DOO\ regions account respectively for 1.7, 0.4, 0.1, 0.006, and 0.003\%
of the total. There are neither \DDD\ (permanently
disconnected field lines) nor \DDO\ (logically equivalent
to the \ODD\ at $R_\sun$) areas on the computed maps.

{
By calculating magnetic fluxes in different areas of these maps, one can quantify the magnetic flux  transfer {in our evolving configuration}  due to various reconnection processes.}
Table \ref{tab:flux} presents for both boundaries all the fluxes of opposite polarities whose calculation does not involve the use of the backward-in-time advection of the footpoints at the upper boundary.
The comparison of the computed flux values indicates that the interchange reconnection with the \CCO, \COO, and \OCC\ types of connectivity changes is a {\hB dominant} process in the magnetic field evolution under study.
The corresponding fluxes are computed for the mesh {\hB used} correctly up to two significant digits.
Naturally, the accuracy decreases in the cases with smaller values of the fluxes, but it still provides consistent results.

%%%%%%%%%%%%%%%%%%%% Usigned fluxes %%%%%%%%%%%%%%%%%%
%
%
%\begin{deluxetable}{c|c|c}
%\tabletypesize{\scriptsize}
%\tablecolumns{2}
%%\tablewidth{0pt}
%\tablecaption{\label{tab:flux} Magnetic Flux Transfer}
%\tablehead{
%\colhead{Boundary} & \colhead{Region} & \colhead{Magnetic Flux, Mx}
%}
%\startdata
%                 & \OOD  & $1.41\times 10^{18}$ \\
%                 & \OCO  & $6.98\times 10^{19}$ \\
%$r=19\,R_{\sun}$ & \OCC  & $4.08\times 10^{20}$ \\
%                 & \DOO  & $5.59\times 10^{16}$ \\
%                 & \DOC  & $2.24\times 10^{18}$ \\
% \hline
%             & \OOC  & $4.95\times 10^{20}$ \\
%             & \ODO  & $4.84\times 10^{18}$ \\
%$r=R_{\sun}$ & \COO  & $4.92\times 10^{20}$ \\
%             & \COD  & $2.95\times 10^{19}$ \\
%             & \CCO  & $4.16\times 10^{20}$ \\
%\enddata
%%\tablecomments{}
%\end{deluxetable}
%%
%%%%%%%%%%%%%%%%%%%% Signed fluxes %%%%%%%%%%%%%%%%%%
%
\begin{deluxetable}{ccccc}
\tabletypesize{\scriptsize}
\tablecolumns{2}
\tablewidth{0pt}
\tablecaption{\label{tab:flux} Magnetic Flux Transfer}
\tablehead{
\colhead{} & \colhead{} & \multicolumn{2}{c}{Magnetic Flux (Mx)} \\
%\colhead{} & \colhead{} & \multicolumn{3}{c}{(Mx)} \\
\cline{3-4}
\colhead{Boundary} & \colhead{Case} & \colhead{Negative} & \colhead{Positive}
}
\startdata
                 & \OOD   & $-0.41 \times 10^{18}$ & $1.00 \times 10^{18}$\\
                 & \OCO   & $-3.47 \times 10^{19}$ & $3.51 \times 10^{19}$\\
$r=19\,R_{\sun}$ & \OCC   & $-2.35 \times 10^{20}$ & $1.72 \times 10^{20}$\\
                 & \DOO   & $-1.51 \times 10^{16}$ & $4.08 \times 10^{16}$\\
                 & \DOC   & $-1.12 \times 10^{18}$ & $1.12 \times 10^{18}$\\
 \hline
%             & \ODO   & $-1.00 \times 10^{18}$ & $3.84 \times 10^{18}$\\
             & \COO   & $-2.27 \times 10^{20}$ & $2.65 \times 10^{20}$\\
$r=R_{\sun}$ & \COD   & $-0.50 \times 10^{19}$ & $2.45 \times 10^{19}$\\
             & \CCO   & $-2.37 \times 10^{20}$ & $1.79 \times 10^{20}$\\
 \hline
             & \OOD   & $-0.29 \times 10^{18}$ & $1.53 \times 10^{18}$\\
             & \OCO   & $-3.40 \times 10^{19}$ & $3.50 \times 10^{19}$\\
$r=R_{\sun}$ & \OCC   & $-2.34 \times 10^{20}$ & $1.73 \times 10^{20}$\\
             & \COO   & $-2.69 \times 10^{20}$ & $2.25 \times 10^{20}$\\
             & \COD   & $-2.48 \times 10^{19}$ & $0.56 \times 10^{19}$\\
             & \CCO   & $-1.73 \times 10^{20}$ & $2.32 \times 10^{20}$\\
\enddata
\tablecomments{The extended slip-back mapping (see Figure \ref{fig-5steps}(a) and Equations (\ref{Smt}) and (\ref{tPt1})) was used to calculate the fluxes at both upper and lower boundaries (top and middle parts of the table), while its dual analog (see Figure \ref{fig-5steps}(b) and Equation (\ref{Smtd})) was used to calculate the fluxes only at the lower boundary (bottom part of the table).}
\end{deluxetable}

\subsection{\hB Evaluation of Footpoint Slippages}
\label{subsec-slippages}
{\hB
If slippage is not due to magnetic reconnection, it is due to:
\begin{enumerate}
\item  Inaccuracies of the field-line tracings. We use a second order,
       predictor-corrector scheme with adaptive
       integration step and have checked that errors are negligible
       compared to diffusion effects by tracing field lines from the
       lower to upper boundary and back. Errors in closed-field regions
       are even smaller.
\item 
{\mybold Inaccuracies in calculating slippage  may also
originate from errors associated with  $\vfp$. While 
 at $R_0$  we use the
analytical expression in Equation (\ref{eq-diff_rot}) to evaluate  $\vfp$,
at $R_1$ we have to rely on the linear interpolation in time of the
 fields obtained from the MHD simulation.  The latter provided
51, equally spaced, field samples  along the interval $[t_0,t_1]$ 
(i.e., $\Delta t=\frac{t_1-t_0}{50}$). Except for locations along
the current sheet, $|\Delta v_{\mathrm{FP}}|/ v_{\mathrm{FP}} \lesssim 2 \%$
between successive time instances. Since at $R_1$   $\vfp$
is mainly due to expanding loops,  only backward advection moves the 
footpoints towards  the current sheet and should be avoided. 
Fortunately, the largest \OCC\ and \OCO\ areas at $R_1$ are calculated by using backward advection only  at the lower boundary. Therefore, the leading type of reconnection is determined correctly.
As clear from Table~\ref{tab:flux}, there are small \DOO\ and \DOC\ areas, whose calculation 
still requires the use of the backward advection at the upper boundary.  
Although these areas are not visible in Figures \ref{R0_R1_SB_maps}a and \ref{R0_R1_SB_i_maps}a, they (and hence their fluxes in Table~\ref{tab:flux}) are 
likely calculated with significant errors.  However, since the disconnected-flux regions are small, the inaccurate calculation of the \DOO\ and \DOC\
areas is not that important.
 Generally speaking, the small values of  
$|\Delta v_{\mathrm{FP}}|/ v_{\mathrm{FP}} $ 
justifies the use of linear interpolation, as
the  error is bounded by $\frac{1}{8}|\vfp''| \Delta t^2$,
with $\Delta t$ very small and
$\vfp$  almost constant. The same conditions should be
verified case by case to ensure the general accuracy of our
method for determining the connectivity changes whose identification involves footpoint advection at $R_1$ only forward in time. }
\item Resistive diffusion. As we show in the introduction,
       this is not an error but rather an inevitable limitation of numerical
       modeling, as it is not possible to model MHD configurations with magnetic Reynolds numbers that
are realistic for the solar corona.
\item Numerical diffusion. Although this is a real error, it is difficult
       to separate it from (3). However, this is in principle possible by
       comparing the connectivity changes with those deduced from the
       voltage difference that is induced by $E_\|$. We postpone work
       on this subject for a future publication, as it is outside
       the scope of the present one.
\end{enumerate}

We now look at the possibility that slippage may be due to magnetic 
reconnection.
}
To verify whether the \CCC\ and \OOO\ regions on our maps are really free of multiple reconnection events, the distribution of footpoint slippages, which are the angular distances between pairs of points $\pP$ and $\tP$, are also computed at {time} $t_1$ {at} both upper and lower boundaries (see panels (a) and (b), respectively, in Figure \ref{fig-fp_slip}).
The obtained distributions have no apparent discontinuities, except for a minor streak extending from $122\arcdeg$ to $145\arcdeg$ in longitude along the boundary of the southern coronal hole.
Thus, there was no reconnection almost anywhere in the \CCC\ and \OOO\ regions during the time interval under study.
A noticeable but smoothly distributed slippage is still present in these regions, which is likely an indication of a significant numerical diffusion caused by the use of a relatively low spatial resolution in our simulation.
\subsection{\hB Analysis of a Field-Line Structure}
Let us consider now an example of the field-line structure of the lateral boundary for one of the reconnected flux tubes determined {by} our method.
As such, we take the open flux tube whose upper-boundary footprint is a tongue-like \OCC\ area attached to one of the two bulges of the PIL (see Figure \ref{R0_R1_SB_maps}(a)).
Its closed lateral boundary is a union of two different magnetic surfaces shown in Figure \ref{fig-RFs}.
One of these surfaces consists of the field lines that thread, among others, those plasma elements which passed at time $t_0$ through the corresponding reconnection site.
The second surface comprises the field lines that thread, among others, those plasma elements which are currently passing the reconnection site.
The former and latter magnetic surfaces are called, respectively, {the} past and present reconnection fronts \citep[RFs,][]{2009ApJ...693.1029T}.

Our past RF is a simple surface formed by open field lines (pink tubes in Figure \ref{fig-RFs}), which become noticeably curved near the heliospheric current sheet.
From comparison of the $\slog Q$ and slip-back maps (see Figures \ref{fig-br_ch_q}(e)--(h) and  \ref{R0_R1_SB_maps}), one can see that this past RF envelops a so-called open hyperbolic flux tube \citep[HFT;][]{2007ApJ...660..863T}, which is a combination of two self-intersecting QSLs.
Here our HFT is apparently a sort of tunnel through which a newly reconnected magnetic flux {fills} the surrounding space in the corona.

Our present RF has a more complex structure, which is actually a part of two intersecting separatrix surfaces that belong to the boundary of the southern coronal hole and that is actively involved {in} the reconnection process at time $t_1$.
One of these separatrix surfaces is woven with closed and disconnected field lines (yellow and gray tubes, respectively, in Figure \ref{fig-RFs}), while the other separatrix surface consists of only open field lines (cyan tubes).
Due to a strong current density in the heliospheric current sheet {that encloses these two surfaces, the latter} intersect by nearly osculating each other along a so-called generalized separator field line \citep[red tube in Figure \ref{fig-RFs}; see also][]{2017ApJ...845..141T}.
{This separator} is likely the site where interchange reconnection takes place, {leading to} the appearance of the \OCC\ regions at both upper and lower boundaries.
{Indeed,} all the {types (\textbf{O}, \textbf{C}, and \textbf{D}}) of {} field lines are present in the neighborhood of the separator.
{Moreover, it evidently} passes through {a strong-current density region,
which implies a locally enhanced resistive slippage of plasma elements from the field lines that are approaching the separator.
As a matter of fact, this slippage ultimately leads to the global effect of the connectivity changes, which is one of the major manifestations of the reconnection process.}

To recover the structure of the present RF, we applied {here} a new technique for determining (quasi)-separatrix surfaces that is based on the use of so-called bracketing field-line pairs \citep{2017ApJ...845..141T}.
In {the} present {work}, the latter are simply pairs of open/closed or open/disconnected field lines whose launching footpoints are ``maximally close'' to each other---the corresponding $\theta$ or $\phi$ coordinates of the footpoints differ from each other by no more than one in their last significant digits.
We find that for a qualitative analysis of the magnetic topology this method works perfectly {well} and, in combination with the slip-back maps, shows a {\hB considerable} heuristic power. 

%
%=======================================================
\begin{figure*}[t]
\centering
\includegraphics[width=0.8\textwidth]{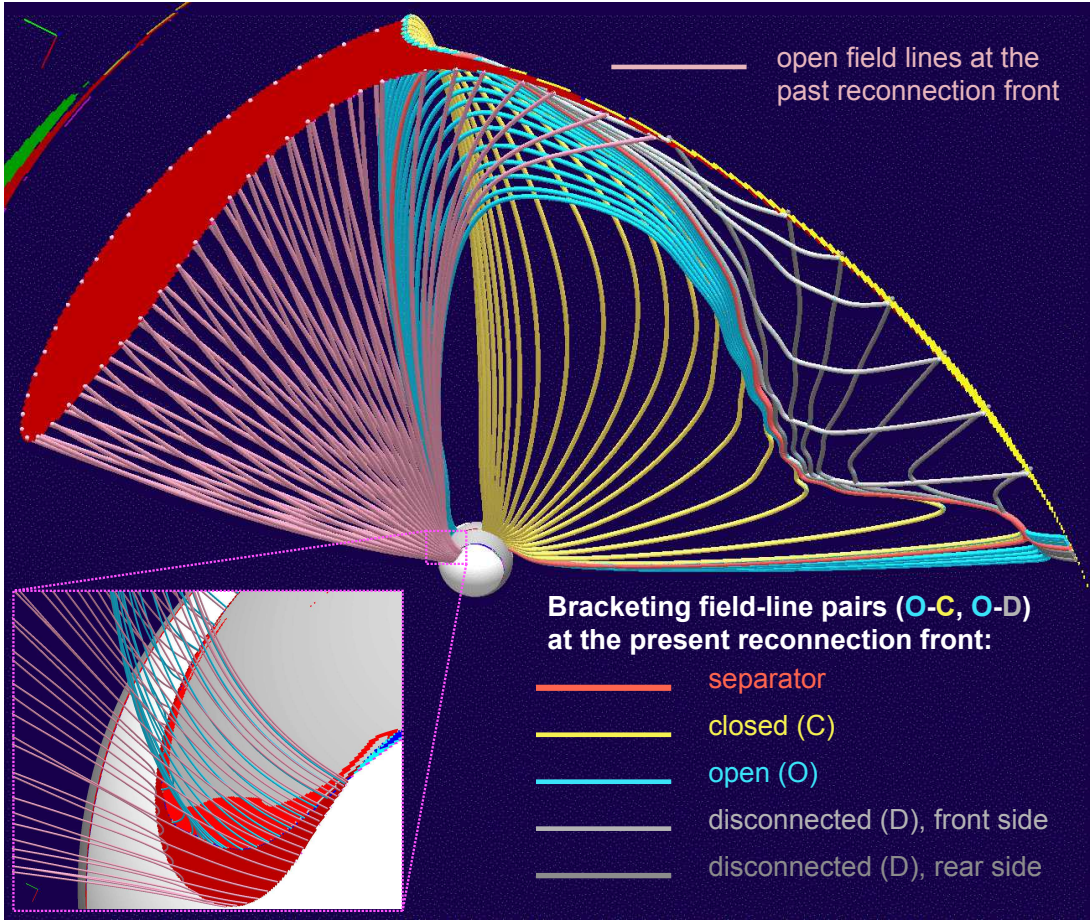}
\caption
{The field-line structure of the boundary of a reconnected open flux tube formed at time $t_1$ by interchange reconnection during the time interval $[t_0,t_1]$; its upper-boundary footprint coincides with the tongue-like \OCC\ area that joins to one of the two bulges of the helmet streamer.
This boundary consists of the past and present reconnection fronts, which are magnetic surfaces, whose field lines thread plasma elements passing through the reconnection site at time $t_0$ and $t_1$, respectively.
The reconnection site here is a neighborhood of the separator field line (red tube), where all three types of closed, open, and disconnected field lines approach one another and reconnect due to a locally enhanced parallel electric field proportional to the current density.
The field-line structure of the present reconnection front is recovered with the help of a technique based on the use of so-called bracketing field-line pairs \citep{2017ApJ...845..141T}.
}
\label{fig-RFs}
\end{figure*}
%%%%%%%%%%%%%%%%%%%%%%

%=======================================================
%\begin{figure}[t]
%\centering
%\includegraphics[width=0.477\textwidth]{reconnection}
%\caption
%{Five examples of magnetic reconnection (\CCO, \COD, \COO, \ODO, \OOC,
%according to the classification of Figure~\ref{fig-classification}) occurring
%in  the simulation of \citet{2005ApJ...625..463L}.
%The labeled field lines corresponds to the cartoons in 
%Figure~\ref{fig-classification} and are calculated through
%the slip-back method. The black field lines are determined
%heuristically as approximate renditions of the field lines
%with which reconnection occurs.
%}
%\label{fig-reconnection}
%\end{figure}
%%%%%%%%%%%%%%%%%%%%%%

%%%%%%%%%%%%%%%%%%%%%%
\section{DISCUSSION AND CONCLUSION}
\label{sec-dis}
{We have developed a method for 
tracking topological changes of the field-line connectivity in evolving magnetic configurations.
The method is based on the computation of a so-called slip-back mapping by using the time sequences of magnetic and velocity fields as input data.
For each mesh point of the lower and upper spherical boundaries and a given time interval $[t_0,t_1]$, we compute the image of this point due to the slip-back mapping.
In turn, the latter is extended by tracing a field line from the image point to obtain the five-step mapping, which is composed of alternating three field-line mappings (at time $t_1$, $t_0$, and $t_1$) and two footpoint-advection mappings (one backward and the other forward in time).
In this way, we associate the mesh point with the sequence of three field lines, the first and third of which refer to the current time $t_1$, while the second to the past time $t_0$.
We represent each sequence by a code word composed of three letters that designate the connectivity types of the field lines in the sequence: \textbf{O} (open), \textbf{C} (closed), and \textbf{D} (disconnected).
We assign also a color code to each of these code words and respective mesh point.

This provides us with a color-coded visualization of the connectivity changes that Lagrangian plasma elements have undergone during the time interval $[t_0,t_1]$ before reaching the mesh points at time $t_1$.
We call this visualization the ``slip-back map'', which is in essence a map of footprints of those reconnected flux tubes whose field lines changed the type of their connectivity to the spherical boundaries in various ways.
The \CCC, \DDD, and \OOO\ regions in our maps are the exception, since they effectively signify no changes of the connectivity type.
They appear as ``trivial'' cases in our classification, although this does not necessarily exclude any reconnection in these regions, because this process might also occur between the field lines that remain connected to the same boundaries.

In contrast, the reconnection processes that do lead to changes of the connectivity type are well identified by the method, which reveals, in particular, for the lower boundary the following cases: \ODO\ (opening up of closed field lines), \COD\ (closing down of open field lines), \CCO, \COO, \ODD, and \OOC\ (interchange reconnection).
Similarly, we obtain for the upper boundary: \OCO\ (opening up of closed field lines by both solar-wind advection and reconnection), \DOC\ (closing down of open field lines), \DDO, \DOO, \OCC, and \OOD\ (interchange reconnection).

To identify the counterpart or conjugate footprints of reconnected flux tubes, we also constructed and implemented a dual analog of the five-step mapping.
This is essentially the same five-step mapping, except that the mesh point and its image are shifted in it for one step forward.
Naively, one would think that the counterpart footprints can be obtained in a much simpler way by using the field-line mapping of the footprints already determined via the five-step mapping.
However, this approach appears to be insufficiently accurate near the part of the footprint boundary that corresponds to the reconnection front at time $t_1$, which includes (quasi-)separatrix surfaces with highly divergent field lines.
Due to this divergence of the field lines, the counterpart footprint obtained with this method significantly loses accuracy at this place, as compared to what one obtains via the dual five-step mapping.

To facilitate interpretation of the slip-back maps, we have developed two different color palettes for their color coding: 
\begin{enumerate}
\item A ``sharp palette'' makes it easier to discriminate between neighboring regions that correspond to different code words.
It comes in two variants, one for the lower and one for the
upper boundary, so that logically equivalent regions are colored in the same colors of different hues.
Then these regions may easily be identified and distinguished from each other if they appear on the same map, for example, due to the dual five-step mapping.  
The sharp palette is applied in Fig.~\ref{R0_R1_SB_maps}.
\item An ``interchange palette", as employed 
in Fig.~\ref{R0_R1_SB_i_maps}, is intended to help quickly identify regions that represent interchange reconnection.
It comes also in two variants, one for the lower and one for the upper boundary, for the same reason as for the sharp palette.
\end{enumerate}

The present implementation of our method implies the detection of the connectivity changes via tracking the footpoint displacements, which is unfortunately not always possible.
However, we have demonstrated that the method admits an extension that is free of this limitation and that will allow one to analyze fairly general MHD evolutions defined in terms of the time sequences of magnetic and velocity fields.

In spite of this limitation,
we have successfully applied the present version of the method to an idealized model of the solar corona driven by differential solar rotation.
In particular, we have identified and classified a number of  reconnection events occurring in such a global configuration.
For one of the prominent \OCC\ regions, we have analyzed the field-line structure of the corresponding reconnected flux system, including the site where the interchange reconnection takes place.
This reconnection site appears to be a generalized separator field line that lies in the helmet streamer at the intersection of separatrix surfaces dividing the configuration into closed, open, and disconnected flux systems.
The {\hB illustration} demonstrates {\hB considerable} heuristic power of the technique we developed, which can shed light on the different magnetic reconfiguration processes occurring in the corona.
}

In particular, 
{as the} method has a natural capability to identify regions of interchange reconnection,
we anticipate that, in conjunction with MHD modeling, it will be useful for the analysis of solar wind measurements
from the Solar Probe Plus and Solar Orbiter missions.
We plan to make the {proposed} generalization of {our} method and apply it for testing the hypothesis that highly variable solar wind originates at interchange reconnection sites.
%%%%%%%%%%%%%%%%%%%%%%
%%%%%%%%%%%%%%%%%%%%%%
%
%
\acknowledgments

This research was supported by NASA's HSR,  and HGI programs, LWS grant
80NSSC20K0192,
NSF grant AGS-1560411,
and AFOSR contract FA9550-15-C-0001.
Computational resources were provided by NSF's XSEDE and NASA's NAS.

\bibliographystyle{apj}
\bibliography{mybib}

\end{document}